\documentclass[conference]{IEEEtran}

\usepackage{cite}
\usepackage{amsmath,amssymb,amsfonts}
\usepackage{graphicx}
\usepackage{xcolor}

\newcommand{\figref}[1]{Fig.~\ref{#1}}   

\newcommand{\mH}{\mathbf{H}}
\newcommand{\mX}{\mathbf{X}}
\newcommand{\mY}{\mathbf{Y}}
\newcommand{\mZ}{\mathbf{Z}}
\newcommand{\va}{\mathbf{a}}

\newcommand{\vp}{\mathbf{p}}
\newcommand{\fUL}{f_{\mathrm{UL}}}
\newcommand{\fDL}{f_{\mathrm{DL}}}
\newcommand{\Kp}{K_{\mathrm{p}}}

\newcommand{\UL}{\mathrm{UL}}
\newcommand{\vg}{\mathbf{g}}
\newcommand{\vy}{\mathbf{y}}
\newcommand{\vn}{\mathbf{n}}

\newcommand{\vz}{\mathbf{z}}
\newcommand{\vnu}{\boldsymbol{\nu}}
\newcommand{\vmu}{\boldsymbol{\mu}}
\newcommand{\vdelta}{\boldsymbol{\delta}}
\newcommand{\mA}{\mathbf{A}}
\newcommand{\mC}{\mathbf{C}}
\newcommand{\mD}{\mathbf{D}}
\newcommand{\mJ}{\mathbf{J}}
\newcommand{\mW}{\mathbf{W}}
\newcommand{\mGamma}{\boldsymbol{\Gamma}}

\usepackage[hidelinks]{hyperref}

\begin{document}

\title{GIST: Location-Dependent Uplink Path Structure Transfer for FDD Downlink Channel Estimation}

\author{\IEEEauthorblockN{Jinghan Zhang\IEEEauthorrefmark{1}\IEEEauthorrefmark{2},
Xitao Gong\IEEEauthorrefmark{1},
Qi Wang\IEEEauthorrefmark{1},
Mate Boban\IEEEauthorrefmark{1},
and Giuseppe Caire\IEEEauthorrefmark{2}}
\IEEEauthorblockA{\IEEEauthorrefmark{1}Huawei Heisenberg Research Center, Munich, Germany}
\IEEEauthorblockA{\IEEEauthorrefmark{2}Department of Electrical Engineering and Computer Science, Technical University of Berlin, Berlin, Germany}}

\maketitle

\begin{abstract}
A key challenge of downlink (DL) channel estimation in frequency division duplex (FDD) massive multiple-input multiple-output (MIMO) systems is the pilot overhead,
which grows with the numbers of antennas and subcarriers.
However, a prior on path delays and directions shared by the uplink (UL) and DL channels can reduce this overhead.
A model trained only on UL channels measured at a base station (BS) can predict the path structure (delays,
directions and power gains) from the location of a user equipment (UE) and thereby assist DL channel estimation.
In this paper, we propose Gaussian-Informed Structure Transfer (GIST), in which each Gaussian primitive models one propagation path,
whose delay, directions and power gain are predicted from the UE location relative to the primitive center.
The complex gains of the predicted paths are estimated from DL pilots by linear minimum mean square error estimation with this prior.
Experiments on a ray-traced and a measured dataset show that GIST outperforms all baselines, reaching a lower DL estimation error with fewer pilot subcarriers.
Further experiments demonstrate its robustness to UE location errors, its low storage and computational requirements.
\end{abstract}

\section{Introduction}\label{sec:intro}

In frequency division duplex (FDD) systems, the uplink (UL) and downlink (DL) occupy different carrier frequencies, separated by more than the coherence bandwidth.
Since the phase of each propagation path depends on the carrier frequency,
instantaneous reciprocity is lost and the base station (BS) cannot directly reuse its UL channel estimate as the DL channel.
Especially in FDD massive multiple-input multiple-output (MIMO) systems, the pilot overhead is substantial,
as conventional estimators such as least squares require DL pilots scaling with the number of BS antennas~\cite{khalilsarai2019covariance,han2019reconstruction}.
Nevertheless, UL and DL signals propagate along the same paths, sharing frequency-independent delays and directions,
a property known as partial reciprocity~\cite{zhong2020reciprocity,han2019reconstruction}.
Delays and directions from the UL can thus assist DL estimation and reduce the pilot overhead.

\begin{figure}[!t]
  \centering
  \vspace*{2.5pt}
  \includegraphics{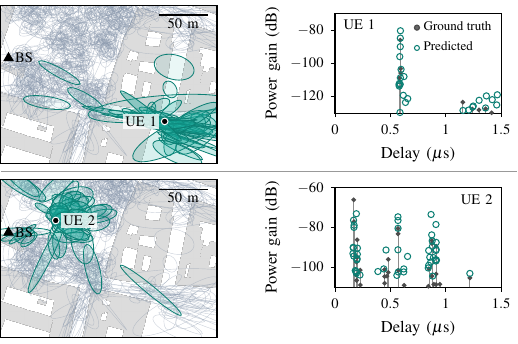}
  \caption{Activated Gaussian primitives (left) and DL paths (right) at two UE locations.
  Filled primitives with the largest UL power gains form the predicted paths.}
  \label{fig:primitives}
\end{figure}

\begin{figure*}[!t]
  \centering
  \includegraphics[width=\textwidth]{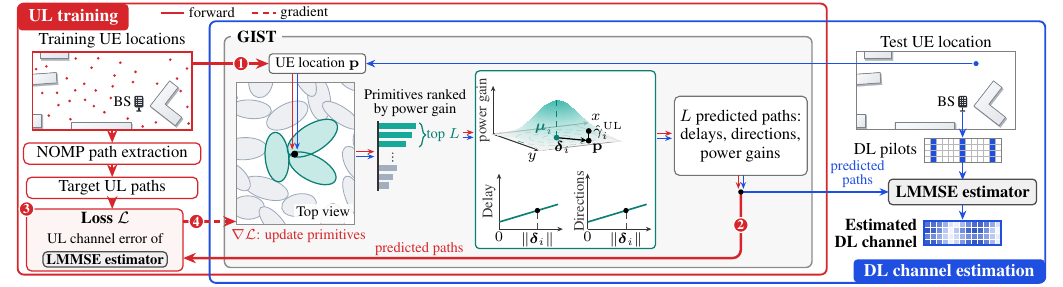}
  \caption{\textbf{Overall process of GIST.} In UL training (red), the primitives are trained on target UL paths.
  In DL channel estimation (blue), the paths predicted at the UE location form the LMMSE prior.
  Each primitive models one path, whose power gain is a Gaussian function of the UE location centered at the primitive,
  whereas its delay and directions are approximated as linear functions of this location.}
  \label{fig:overview}
\end{figure*}

An intuitive approach to avoid the pilot overhead is to predict the DL channel directly from the UL channel~\cite{liu2021fire,zhao2023nerf2,wen2026wrfgs}
or the location~\cite{chatelier2025location,umer2026ngrf} of the user equipment (UE).
However, the instantaneous DL channel predicted without DL pilots may contain phase errors and deviations caused by small
environmental changes or UE location errors.
Another way is to obtain the path delays and directions from UL channels as the prior of
pilot-based DL estimation, as in parametric reconstruction methods~\cite{han2019reconstruction,ji2021multipath}.
Per-UE path extraction at runtime in these methods, however, can challenge real-time operation.
In a covariance-based method proposed in~\cite{khalilsarai2019covariance}, the channel is estimated from the DL pilots along a few beam directions
selected from a UL covariance extrapolated to the DL, while the path delays remain unknown.
In addition, these methods still rely on recent UL sounding of each UE.

In contrast, the path structure (delays, directions and power gains) predicted from the UE location as in \figref{fig:primitives} can assist DL estimation
without such sounding, with such a prediction relying on site-specific information.
This information is provided by a 3D environment model or channels measured at training locations.
The site-specific prior is obtained from an environment model by ray tracing~\cite{delmoro2025digitaltwin}.
A site-specific distribution of path parameters can be learned from measured channels~\cite{bock2025physics},
and a channel knowledge map (CKM) built from these channels stores per-cell UL path structures for UL estimation at the BS~\cite{jiang2025ckm}.
The DL priors of diffusion models~\cite{fesl2024diffusion,hu2026eadm} are instead learned from DL training channels.

These site-specific priors depend on an accurate environment model~\cite{delmoro2025digitaltwin}, ignore the UE location~\cite{bock2025physics},
require storage that grows with the number of grid cells~\cite{jiang2025ckm} or DL training channels~\cite{fesl2024diffusion,hu2026eadm}.
Recent Gaussian splatting-based methods model wireless channels~\cite{wen2026wrfgs,umer2026ngrf,bjorndahl2026beammix,%
zhang2026bisplat,zhou2025bwgs} with discrete Gaussian primitives.
Based on this concept, we propose Gaussian-Informed Structure Transfer (GIST),
in which each Gaussian primitive on the plane of UE locations models one propagation path present over a region of locations, as shown in \figref{fig:overview}.
The paths predicted at the UE location serve as the prior of DL estimation from DL pilots.
The contributions are as follows:
\begin{itemize}
\item We propose GIST, a model that is learned only from UL channels at training locations and can provide a location-dependent prior for DL channel estimation.
\item GIST requires less storage than the compared baselines with a negligible increase in latency.
\item With few DL pilot subcarriers, GIST is more accurate than all baselines on ray-traced and measured channels.
\end{itemize}

\section{System Model}\label{sec:system}

An FDD MIMO link with orthogonal frequency division multiplexing (OFDM) connects BS and UE uniform rectangular arrays of $M$ and $N$ antennas, respectively.
The UL band and the DL band, with carrier frequencies $\fUL$ and $\fDL$, each contain $K$ OFDM subcarriers at the offsets $f_k$ from its carrier.
The DL channel $\mH_k\in\mathbb C^{N\times M}$ on subcarrier $k$ is modeled as a superposition of propagation paths,
\begin{equation}\label{eq:channel}
\mH_k=\sum_{\ell} g_\ell\, e^{-j2\pi f_k\tau_\ell}\,\va_{\mathrm{UE},\ell}\,\va_{\mathrm{BS},\ell}^{\mathsf T},
\end{equation}
with the complex gain $g_\ell$, the delay $\tau_\ell$ and the complex array responses $\va_{\mathrm{BS},\ell}$, $\va_{\mathrm{UE},\ell}$ at $\fDL$,
which depend on the DL angle of departure and angle of arrival of path $\ell$,
given as unit vectors from the two arrays along the path.

The DL channel and the UL channel share the path delays and directions~\cite{zhong2020reciprocity}, and the power gains $\gamma_\ell=|g_\ell|^2$ change only slightly
between the carriers~\cite{han2019reconstruction}.
The UL channel at a location can therefore reflect the DL path structure (delays, directions and power gains) at the same location.
Since the phases of the complex gains change with the carrier frequency, the DL complex gains are estimated from DL pilots.

The BS sends $T$ pilot symbols on each of $\Kp$ equispaced pilot subcarriers, and the UE receives on pilot subcarrier $k$
\begin{equation}\label{eq:pilot}
\mY_k=\mH_k\mX_k+\mZ_k,
\end{equation}
with the random pilot matrix $\mX_k\in\mathbb{C}^{M\times T}$ of independent entries of magnitude $1/\sqrt{M}$ and the noise $\mZ_k$ of
independent $\mathcal{CN}(0,\sigma^2)$ entries.
The reference signal-to-noise ratio (SNR) is the mean power per entry of the UL channel at the UE location divided by $\sigma^2$,
where each OFDM resource element (RE) carries unit transmit energy.
Each pilot symbol occupies one RE, giving a pilot overhead of $\Kp T$ REs.
Since $T<M$, the $NT$ observations in $\mY_k$ are fewer than the $NM$ unknowns in $\mH_k$.
The path structure could reduce the $KNM$ unknowns in all $\mH_k$ to one complex gain per path.

Given the UE location $\vp$ from positioning~\cite{3gpp2021positioning},
$\mH_k$ on all $K$ subcarriers is estimated as $\hat{\mH}_k$ from $\mY_k$ and $\mX_k$ on the pilot subcarriers with a path structure prior at $\vp$
learned from UL channels that the BS measures at known training locations in a static environment.
The accuracy of $\hat{\mH}_k$ is measured by the normalized mean square error (NMSE) per link,
$\mathrm{NMSE}=\sum_{k}\|\hat{\mH}_k-\mH_k\|_{\mathrm{F}}^2/\sum_{k}\|\mH_k\|_{\mathrm{F}}^2$.
We aim to minimize this NMSE for a given pilot overhead.

\section{Proposed Method}\label{sec:method}

\subsection{Overview}\label{sec:overview}

\figref{fig:overview} shows the overall process of GIST,
in which the path structure is predicted by Gaussian primitives learned only from UL channels at sparse training locations.
The $L$ primitives with the largest UL power gains at the UE location $\vp$ form $L$ predicted paths (Section~\ref{sec:prediction}).
The complex gains of these paths are estimated from the DL pilots by the linear minimum mean square error (LMMSE) estimator
with the predicted power gains scaled to the DL carrier as prior variances, and the DL channel is reconstructed on all subcarriers.
The primitives are trained by simulating this estimation on the UL training channels
and minimizing its expected error with target paths extracted from these channels.

\subsection{Gaussian primitives}\label{sec:primitives}

In contrast to the Gaussian splatting-based methods~\cite{wen2026wrfgs,umer2026ngrf,bjorndahl2026beammix,zhang2026bisplat,zhou2025bwgs}
that synthesize the received signal, the $G$ 2D Gaussian primitives in GIST are placed on the horizontal plane of UE locations,
and each primitive models the delay, directions and power gain of one potential propagation path as functions of the UE location.
The responses of both arrays to a path depend on its direction only through the two direction cosines $(u_1,u_2)$
of this direction along the array axes, with $u_1^2+u_2^2\le1$.
The delay and the direction cosines at both arrays form the five path parameters.

The power gain of primitive $i$ is defined by a center $\vmu_i\in\mathbb R^2$, a covariance matrix $\mC_i\in\mathbb R^{2\times2}$ and the peak log power gain $b_i$.
The predicted UL power gain of primitive $i$, evaluated at the UE location $\vp\in\mathbb R^2$ on this plane, is given by
\begin{equation}\label{eq:power}
\hat\gamma_i^{\UL}(\vp)=\exp\!\big(b_i-\tfrac12\vdelta_i^{\mathsf T}\mC_i^{-1}\vdelta_i\big),\quad\vdelta_i=\vp-\vmu_i,
\end{equation}
This power gain decays away from $\vmu_i$ over an extent determined by $\mC_i$, similar to the visibility region of a cluster~\cite{liu2012cost2100}.
The path parameters of primitive $i$ at $\vp$ form the vector $\vz_i(\vp)$, which is approximated as linear in the location
within the extent of the primitive, a region of significant power gain over which the path parameters vary smoothly,
\begin{equation}\label{eq:affine}
\vz_i(\vp)=\vnu_i+\mJ_i\vdelta_i,
\end{equation}
with the path parameters $\vnu_i\in\mathbb R^5$ at $\vmu_i$ and their slopes $\mJ_i\in\mathbb R^{5\times2}$ with respect to the location.

\subsection{Path prediction and DL channel estimation}\label{sec:prediction}

For the $L$ strongest primitives at $\vp$, $L$ paths $\ell=1,\ldots,L$ are predicted with the UL power gains $\hat\gamma_\ell^{\UL}$ from \eqref{eq:power},
and with the delays $\hat\tau_\ell$ and direction cosines from \eqref{eq:affine},
where each delay is clipped to $[0,\tau_{\max}]$ for a delay limit $\tau_{\max}$
and each pair of direction cosines with $u_1^2+u_2^2>1$ is rescaled to the unit circle.
The free-space power gain along a path between two antenna elements is proportional to the squared wavelength,
which changes by the factor $(\fUL/\fDL)^2$ from the UL carrier to the DL carrier.
The UL power gains of all predicted paths are scaled by this factor to the DL power gains $\hat\gamma_\ell=(\fUL/\fDL)^2\hat\gamma_\ell^{\UL}$,
while the delays and directions are reused at $\fDL$.

Let $\hat\va_{\mathrm{BS},\ell}$ and $\hat\va_{\mathrm{UE},\ell}$ be the array responses of \eqref{eq:channel}
at $\fDL$ for the predicted direction cosines of path $\ell$.
The DL channel is modeled as $\mH_k=\sum_\ell g_\ell\mD_{k,\ell}$ with the unknown complex gains $g_\ell$ and the unit-gain path responses
$\mD_{k,\ell}=e^{-j2\pi f_k\hat\tau_\ell}\hat\va_{\mathrm{UE},\ell}\hat\va_{\mathrm{BS},\ell}^{\mathsf T}$.
The observations in \eqref{eq:pilot} on the pilot subcarriers $k_1,\ldots,k_{\Kp}$ take the form $\mY_k=\sum_\ell g_\ell\mD_{k,\ell}\mX_k+\mZ_k$.
Stacked over these subcarriers, the observations form the linear model $\vy=\mA\vg+\vn$ with $\vg=[g_1,\ldots,g_L]^{\mathsf T}$
and $\vn=[\operatorname{vec}(\mZ_{k_1})^{\mathsf T},\ldots,\operatorname{vec}(\mZ_{k_{\Kp}})^{\mathsf T}]^{\mathsf T}$,
where $\operatorname{vec}(\cdot)$ stacks the columns of a matrix.
With the columns of the pilot response matrix $\mA$ written out, this model can be expressed as
\begin{equation}\label{eq:stack}
\underbrace{\begin{bmatrix}\operatorname{vec}(\mY_{k_1})\\\vdots\\\operatorname{vec}(\mY_{k_{\Kp}})\end{bmatrix}}_{\vy}
=\sum_{\ell=1}^{L}g_\ell
\underbrace{\begin{bmatrix}\operatorname{vec}(\mD_{k_1,\ell}\mX_{k_1})\\\vdots\\\operatorname{vec}(\mD_{k_{\Kp},\ell}\mX_{k_{\Kp}})\end{bmatrix}}
_{\text{column }\ell\text{ of }\mA}
+\vn.
\end{equation}
The gains with uniformly random phases follow the zero-mean prior $\vg\sim\mathcal{CN}(\mathbf 0,\mGamma)$
with $\mGamma=\operatorname{diag}(\hat\gamma_1,\ldots,\hat\gamma_{L})$, and their LMMSE estimate $\hat\vg=\mW\vy$ is obtained with
\begin{equation}\label{eq:lmmse}
\mW=\big(\mA^{\mathsf H}\mA+\sigma^2\mGamma^{-1}\big)^{-1}\mA^{\mathsf H}.
\end{equation}
The DL channel on all subcarriers is then estimated as $\hat\mH_k=\sum_{\ell}\hat g_\ell\mD_{k,\ell}$.

Additional paths are extracted from the same DL pilots by Newtonized orthogonal matching pursuit
(NOMP)~\cite{mamandipoor2016nomp} in the variant GIST + NOMP.
Paths with delays up to $\tau_{\max}$ can be extracted without aliasing only when the pilot spacing is below $1/\tau_{\max}$.
The extracted paths join the predicted paths in \eqref{eq:lmmse} at such spacings with estimated power gains as prior variances,
and otherwise only the predicted paths are used.

\subsection{Training}\label{sec:training}

The GIST model contains 21 learnable scalars per primitive: the center $\vmu_i$, the covariance $\mC_i$, the peak log power gain $b_i$,
the path parameters $\vnu_i$ at the center and the slopes $\mJ_i$.
The target paths at each training location are extracted from the UL channel by NOMP.
Each primitive is initialized from a different randomly drawn target path,
with $\vmu_i$ set to its training location, $\vnu_i$ to its delay and direction cosines,
$b_i$ to its log power gain, $\mJ_i$ to zero and $\mC_i$ to a scaled identity matrix.

The predicted paths are useful as the LMMSE prior only to the extent that they lower the DL estimation error.
The loss is therefore built directly from this error.
Since the UL channel shares the path delays and directions with the DL channel,
this error can be evaluated analogously on the UL training channels.
The UL channel at a training location is represented by the target paths with their extracted power gains and independent random phases.
The LMMSE estimate is computed from simulated pilot observations of this UL channel with $\hat\gamma_\ell^{\UL}$ as prior variances.
The unit-gain responses of the predicted paths on all subcarriers form the matrix
\begin{equation}\label{eq:dmat}
\mD=\begin{bmatrix}\operatorname{vec}(\mD_{1,1})&\cdots&\operatorname{vec}(\mD_{1,L})\\\vdots&&\vdots\\
\operatorname{vec}(\mD_{K,1})&\cdots&\operatorname{vec}(\mD_{K,L})\end{bmatrix}.
\end{equation}
Each target path $j=1,\ldots,\bar L$ has the unit-gain response $\bar\mD_{k,j}$ and complex gain $\bar g_j$ with power gain $|\bar g_j|^2=\bar\gamma_j$.
The matrices $\bar\mD$ and $\bar\mA$ follow from \eqref{eq:dmat} and \eqref{eq:stack} with $\bar\mD_{k,j}$ in place of $\mD_{k,\ell}$,
where $\bar\vg=[\bar g_1,\ldots,\bar g_{\bar L}]^{\mathsf T}$ has zero mean under the random phases,
with covariance $\bar\mGamma=\operatorname{diag}(\bar\gamma_1,\ldots,\bar\gamma_{\bar L})$.
With the observations $\vy=\bar\mA\bar\vg+\vn$, the channel estimate $\mD\mW\vy$ on all subcarriers has the error
$(\mD\mW\bar\mA-\bar\mD)\bar\vg+\mD\mW\vn$,
where the first term is caused by the mismatch between predicted and target paths, the second by noise.
Averaged over these phases and noise, the error-to-channel energy ratio can be written as
\begin{equation}\label{eq:risk}
r(\vp)=\frac{\big\|(\mD\mW\bar\mA-\bar\mD)\bar\mGamma^{1/2}\big\|_{\mathrm F}^2+\sigma^2\big\|\mD\mW\big\|_{\mathrm F}^2}
{\big\|\bar\mD\bar\mGamma^{1/2}\big\|_{\mathrm F}^2}.
\end{equation}
In each update, $r(\vp)$ is computed with new random pilot matrices for the next pair of $\Kp$ and reference SNR from a fixed grid.
The loss over a training batch $\mathcal B$ is defined as
\begin{equation}\label{eq:loss}
\mathcal L=\log\Big(\frac{1}{|\mathcal B|}\sum_{\vp\in\mathcal B}r(\vp)\Big).
\end{equation}
The power gains are learned as LMMSE prior variances by differentiating the loss through \eqref{eq:power}, \eqref{eq:affine} and \eqref{eq:lmmse}.

\section{Experiments and Numerical Results}\label{sec:experiments}

\begin{figure}[!t]
  \centering
  \includegraphics{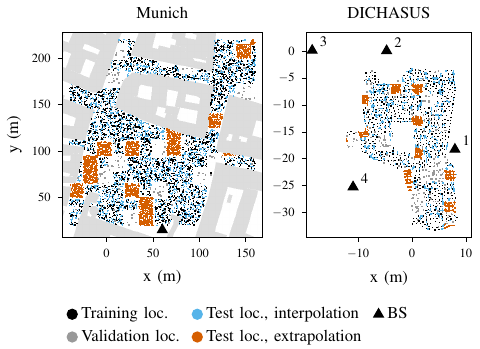}
  \vspace{-4pt}
  \caption{\textbf{Top views of the scenes} with the training, validation and test locations}
  \label{fig:scene}
\end{figure}

\begin{figure}[!t]
  \centering
  \includegraphics{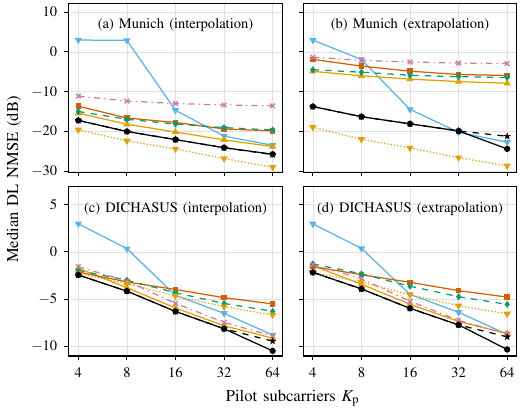}\par\vspace{6pt}
  \includegraphics{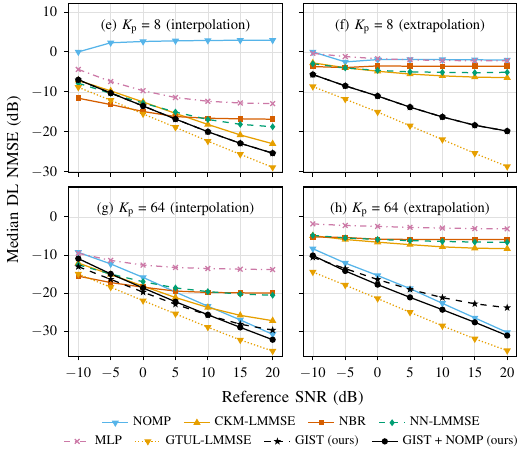}
  \caption{Median DL NMSE versus the number of pilot subcarriers at 10 dB (a)--(d) and versus the reference SNR in Munich (e)--(h),
  with GIST and GIST + NOMP coinciding below $\Kp = 64$}
  \label{fig:main}
\end{figure}

\subsection{Datasets}

The datasets are from ray tracing of the Munich scene in Sionna RT~\cite{hoydis2023sionnart} and from the DICHASUS measurements~\cite{euchner2024dichasus}.
The Munich channels comprise paths with at most two bounces between fixed-orientation $8\times8$ BS and $2\times2$ UE arrays
in bands at $\fUL = 2.55$~GHz and $\fDL = 2.67$~GHz, each with $K = 612$ subcarriers spaced by 30~kHz.
The DICHASUS recordings ad02 and ad04 from the ARENA2036 factory provide a single-antenna UE and four $2\times8$ arrays as separate BSs,
and two measured UL subbands with $K = 376$ subcarriers spaced by 48.9~kHz form a virtual FDD pair with $\fUL = 3.424$~GHz and $\fDL = 3.456$~GHz.

Each scene in \figref{fig:scene} is divided into square blocks of 15~m in Munich and 2~m in DICHASUS.
The validation locations and the extrapolation test locations each occupy 10\% of the blocks, which contain no training locations.
The other 80\% of the blocks contain the interpolation test locations and the training locations,
which lie at least 2~m in Munich and 0.5~m in DICHASUS from the blocks without training locations.
The training set has 2000 samples in Munich and 620 in DICHASUS.
The test set has 2000 interpolation and 2000 extrapolation samples in Munich.
Each DICHASUS array has 3750 for interpolation and 5646 for extrapolation.

\subsection{Experimental setup}

The three compared baselines use UL paths of training samples extracted and scaled as in GIST.
In nearest-neighbor LMMSE (NN-LMMSE), the UL paths of the nearest training location form the LMMSE prior in \eqref{eq:lmmse}.
CKM-LMMSE, adapted from~\cite{jiang2025ckm}, instead takes this prior from the UL paths fitted jointly to the training channels in each grid cell,
whose side length is set to 5~m in Munich and is chosen as 2~m on validation locations in DICHASUS.
In nearest beamformed reconstruction (NBR), adapted from~\cite{han2019reconstruction},
pilots with the same REs and energy as in the other methods are beamformed
along the path directions of the nearest training location, and the gains are estimated by least squares.
The other two reference methods are NOMP using the DL pilots alone and GTUL-LMMSE,
whose prior consists of up to $L$ paths extracted as in GIST from the ground-truth UL (GTUL) channel at the UE location.
GTUL-LMMSE thus corresponds to error-free recent UL sounding of the UE.
To verify the benefit of the primitives, they are replaced by a multilayer perceptron (MLP) trained with the same loss,
with four hidden layers of 256 units.
In both datasets, $G = 1024$ primitives are used, with $L = 48$ paths retained at each UE location in Munich and 96 in DICHASUS.
GIST + NOMP adds NOMP paths only at $\Kp = 64$, where the pilot spacing is below $1/\tau_{\max}$ for the per-array $\tau_{\max}$ of 1.9--2.2~$\mu$s.

The pilots occupy $\Kp\in\{4,8,16,32,64\}$ subcarriers with eight symbols each, and the reference SNR takes values in $\{-10,-5,\dots,20\}$~dB.

\subsection{DL estimation accuracy}

\begin{figure}[!t]
  \centering
  \includegraphics{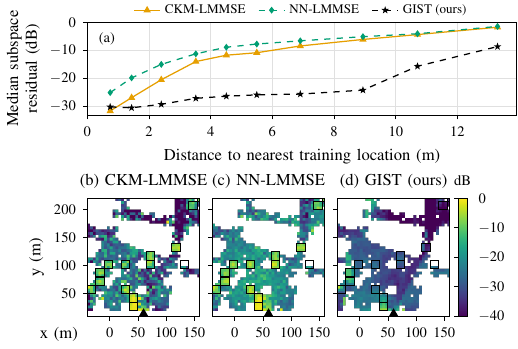}
  \caption{Median subspace residual in Munich (a) versus the distance to the nearest training location and (b)--(d) per $5\times5$~m cell,
  with extrapolation blocks outlined}
  \label{fig:residual}
\end{figure}

\begin{figure}[!t]
  \centering
  \includegraphics{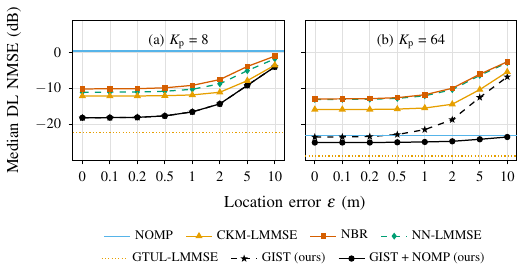}
  \caption{Median DL NMSE in Munich at 10 dB versus the standard deviation of the location error,
  with GIST and GIST + NOMP coinciding at $\Kp = 8$}
  \label{fig:poserr}
\end{figure}

\begin{figure*}[!t]
  \centering
  \includegraphics{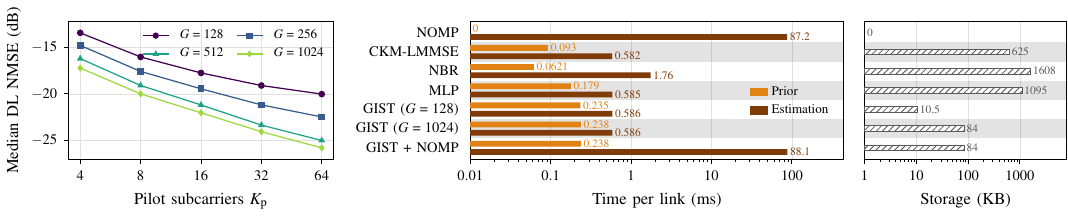}
  \caption{Median DL NMSE with GIST for interpolation in Munich at 10 dB versus the number of pilot subcarriers for different numbers of primitives (left),
  latency per link for obtaining the prior and for the estimation (middle), and storage (right) of each method}
  \label{fig:cost}
\end{figure*}

The median NMSE with GIST in \figref{fig:main}(a)--(d) is lower than those with all baselines at every $\Kp$
and than that with NOMP up to $\Kp = 16$, with large gaps of 9 to 13~dB to the baselines in extrapolation in Munich.
Here, GIST with four pilot subcarriers outperforms the baselines with 64
and reaches a median NMSE of $-15$~dB with eight, whereas NOMP requires 32.
The MLP is much less accurate than GIST in Munich and up to 1.1~dB less accurate on DICHASUS despite over 10 times as many parameters.

GTUL-LMMSE serves as an error lower bound for UL path priors only in Munich,
where the ray-traced UL paths at the UE location are well separated.
However, GIST outperforms GTUL-LMMSE on the measured DICHASUS channels in \figref{fig:main}(c)--(d) at $\Kp \ge 8$.
Some of the paths fitted to the measured UL channel at the UE location may transfer poorly to the DL carrier and weaken the GTUL-LMMSE prior.
Such location-dependent paths may be suppressed by learning each GIST primitive from many training locations.

GIST or GIST + NOMP is the most accurate in extrapolation at every SNR in \figref{fig:main}(e)--(h) apart from GTUL-LMMSE.
NBR, which beamforms the pilots along the paths of the nearest training location, is more accurate than GIST in interpolation only at low SNRs.
As the SNR rises, the GIST error continues to decrease, whereas the NN-LMMSE and NBR errors saturate,
possibly because the paths of the nearest training location differ from the true paths.
GIST + NOMP outperforms both GIST and NOMP at $\Kp = 64$ except at low SNRs, where noisy NOMP paths can degrade GIST + NOMP.
GIST at $\Kp = 8$ is only 2 to 4~dB above GTUL-LMMSE in interpolation without UL sounding of the UE.

\subsection{Accuracy of the predicted path structure}

To examine the source of GIST's accuracy in extrapolation, the predicted path structure at a test location is assessed independently of the DL pilots
by the subspace residual
\begin{equation}\label{eq:residual}
  \frac{\sum_{\ell}\gamma_\ell\,\|\mathbf{d}_\ell-\hat{\mathbf{\Pi}}\mathbf{d}_\ell\|^2}
       {\sum_{\ell}\gamma_\ell\,\|\mathbf{d}_\ell\|^2},
\end{equation}
where $\mathbf{d}_\ell$ is the unit-gain response of ray-traced DL path $\ell$ over all subcarriers and antenna pairs,
and $\hat{\mathbf{\Pi}}$ is the orthogonal projector onto the subspace spanned by the DL responses of the paths in the prior of each method.
The GIST residual in \figref{fig:residual}(a) falls below the NN-LMMSE and CKM-LMMSE residuals except near training locations,
remaining under $-24$~dB up to 9~m from them.
It is below $-10$~dB in 95\% of the extrapolation cells in \figref{fig:residual}(b)--(d), versus 16\% for NN-LMMSE and 38\% for CKM-LMMSE.
Since each primitive models its path over the whole Gaussian extent, paths observed at training locations can carry over to nearby unsampled locations,
whereas the NN-LMMSE and CKM-LMMSE priors stay tied to one training location or grid cell.

\subsection{Robustness to UE location errors}

\figref{fig:poserr} shows the median DL NMSE averaged over interpolation, extrapolation and five draws of random location errors.
Each reported UE location is perturbed by an independent Gaussian error with standard deviation $\epsilon$ per horizontal coordinate,
while the training locations are exact.
GIST outperforms all baselines at every $\epsilon$ and for both $\Kp$,
although its margin over CKM-LMMSE at $\Kp = 8$ shrinks from 6~dB at $\epsilon = 0$ to 0.5~dB at 10~m.
GIST alone deviates by less than 0.5~dB from NOMP at $\Kp = 64$ for $\epsilon$ up to 0.5~m,
which covers the 3GPP Release 17 target of horizontal errors below 1~m for 90\% of commercial UEs ($\epsilon = 0.47$~m)~\cite{3gpp2021positioning},
and GIST + NOMP remains more accurate than NOMP up to $\epsilon = 10$~m.

\subsection{Complexity and storage}

The path prediction is $\mathcal O(G)$ per link, whereas the DL estimation is independent of $G$.
As $G$ increases from 128 to 1024 in \figref{fig:cost}, the median NMSE with GIST falls at every $\Kp$ and the storage grows linearly to 84~KB,
whereas the path prediction latency stays at about 0.24~ms because the primitives are evaluated in parallel on a GPU.
GIST with $G = 1024$ requires far less storage than CKM-LMMSE, the MLP and NBR while GIST's total latency is only 0.15~ms above that of CKM-LMMSE for $\Kp = 64$.
GIST + NOMP adds no storage, but its latency is dominated by the NOMP grid search and is about 100 times that of GIST.
Overall, GIST combines a low latency with a very small storage footprint, which could support on-device deployment and frequent model distribution.
GIST training takes only about 96~s for Munich on a single GPU.

\section{Conclusions}\label{sec:conclusion}

We proposed GIST for FDD DL channel estimation with a compact, location-dependent prior learned only from UL channels at training locations.
Results on ray-traced and measured channels demonstrate accurate estimation with few DL pilot subcarriers,
outperforming the compared baselines.
Although GIST takes the UE location as input, it remains highly robust to location errors.
Its low storage requirements and estimation latency could make GIST a promising candidate for resource-constrained devices and frequent model distribution.
This work considers UE locations on a horizontal plane, a fixed UE array orientation and exact training locations.
Future work could extend GIST to varying UE heights by 3D primitives at the same $\mathcal O(G)$ complexity
and to varying UE orientations by rotating global-frame directions into the UE array.
Collaborative updates using UL observations collected across devices could also adapt the learned prior to environmental changes.

\bibliographystyle{IEEEtran}
\bibliography{references}

\end{document}